# Verifying quantum superpositions at metre scales

D.M. Stamper-Kurn[1,2], G.E. Marti[3] and H. Müller[1]

**While the existence of quantum superpositions of massive particles over microscopic separations has been established since the birth of quantum mechanics, the maintenance of superposition states over macroscopic separations is a subject of modern experimental tests. In Ref. [1], T. Kovachy et al. report on applying optical pulses to place a freely falling Bose-Einstein condensate into a superposition of two trajectories that separate by an impressive distance of 54 cm before being redirected toward one another. When the trajectories overlap, a final optical pulse produces interference with high contrast, but with random phase, between the two wave packets. Contrary to claims made in Ref. [1], we argue that the observed interference is consistent with, but does not prove, that the spatially separated atomic ensembles were in a quantum superposition state. Therefore, the persistence of such superposition states remains experimentally unestablished.**

The authors of Ref. [1] incorrectly equate the observation of interference with the existence of a phase-coherent quantum superposition between the separated atomic samples. The distinction between interference and phase coherence is emphasized by Anderson's hypothetical experiment involving two independently produced, "non-communicating" volumes of superfluid helium. He pointed out that connecting the two volumes by a narrow orifice would result in a Josephson current. The relative phase determined from the Josephson relation would have a random value. No consequent measurement on a single experimental realization of this setup could determine whether this phase had been established before or only after the Josephson current was produced [2].

Such *gedanken* experiments have been realized in the laboratory, demonstrating interference between two independently generated light beams [3], and between two independently produced Bose-Einstein condensates [4]. In both these examples, the spatially separated, indistinguishable quantum objects – photons in one case, sodium atoms in the other – had no defined quantum coherence between them. Each sample could have interacted with its own local environment, and experienced uncorrelated perturbations therefrom. Yet, in each repetition of the experiment the interference was high in contrast, while the phase of the interference was irreproducible between repetitions. The same behavior is observed by Kovachy et al.

Phase-coherent quantum superposition states are characterized by first-order coherence. First-order coherence measures the expectation value of a product of two field operators $\langle \psi_A^\dagger \psi_D \rangle$. In a many-body system, this expectation value appears in the off-diagonal element of the one-body reduced density matrix. Such coherence is measured by a two-slit interference experiment: the quantum fields emanating from two points, $A$ and $D$, are allowed to interfere. The presence of first-order coherence, i.e. of quantum superposition states, is indicated by an interference pattern with a *determinate* phase. While it is possible that the random-phase interference observed by Kovachy et al. is caused only by technical imperfections in their optical pulses, their observation is also consistent with the lack of first-order coherence and of coherent quantum superpositions.

In contrast, second-order coherence measures the expectation value of a product of four field operators, $\langle \psi_A^\dagger \psi_B^\dagger \psi_C \psi_D \rangle$, and is an element within the two-body density matrix. Second-order coherence is indicated by the fact that the interference pattern produced by two quantum fields, say those emanating from points $A$ and $D$, is the same as that that between two other quantum fields, say those emanating from points $B$ and $C$. In the experiment of Kovachy et al., the points $A$ and $B$ correspond to locations within the gas on the upper interferometer path, and the points $C$ and $D$ to positions within the gas on the lower path. Their observation that the interference phase in one portion of the gas is equal to that in another portion of the gas demonstrates the existence of second-order, but not first-order, coherence.

[1] Department of Physics, University of California, Berkeley, CA 94720
[2] Materials Sciences Division, Lawrence Berkeley National Laboratory, Berkeley, CA 94720
[3] JILA, National Institute of Standards and Technology and University of Colorado, Boulder, CO 80309; and Department of Physics, University of Colorado, Boulder, CO 80309

Thus, the experiment of Kovachy et al. does not demonstrate the existence of quantum superposition states of massive particles over metre length scales. The second-order coherence observed in the experiment is immune to perturbations that are common across the sub-millimeter length scale of each of the spatially separated clouds, but that differ arbitrarily over the metre-scale distance between the two paths of their atomic interferometer. Such perturbations can arise from technical imperfections or intrinsic atomic interactions. In addition, and directly relevant to the claims made by Kovachy et al., these perturbations would arise from exotic effects proposed in theories of continuous spontaneous localization or gravitationally induced decoherence [5, 6]. Indeed, these exotic localization effects, if they localize particles with metre-scale resolution and also respect the indistinguishability of identical quantum particles, can even go so far as determining the exact mass, and hence the exact number of atoms in each interferometer path. Such localization would completely eliminate first-order coherence between the two interferometer paths, so that the one-body density matrix becomes that of a mixed state [6]. The state produced by such localization is a "quantum superposition" only insofar as it is composed of identical bosons, whose wavefunction must be symmetric under particle exchange. Yet, as in Anderson's *gedanken* experiment [2], the two atomic wave packets will still show high-contrast interference, with an interference phase that is random between experiments [7, 8]. We point out that our argument invalidates the claim by Nimmrichter and Hornberger [6] that single-shot measurements of atom interferometers serve to test their phenomenological model for the decay of macroscopic quantum superposition states.

The second-order coherence observed by Kovachy et al. does demonstrate that each of the separated Bose-Einstein condensates remains coherent over its sub-millimetre size during the 1 second time of propagation. Matter-wave coherence over similar length and time scales has been observed previously, as summarized in Extended Data Figure 3 and Table 1 of Ref. [1]. In those previous works, the existence of a determinate phase is confirmed by comparing the phases of two well-separated atomic interferometers, allowing for the elimination of common-mode technical noise. The observations of Kovachy et al. do rule out exotic effects that would cause the position of each and every individual atom to be independently measured with metre-scale resolution. However, such effects would violate the principle of the indistinguishability of identical particles, and are thus implausible (as discussed in Ref. [6]).

Verification of a quantum mechanical superposition requires the measurement of a determinate phase in order to distinguish a pure quantum state from a statistical mixture of several pure states. Once information about the phase is lost, whether due to measurement noise, interactions with the environment, or a fundamental source of decoherence, no further measurement can distinguish between a quantum superposition and a mixed state. In the case of Kovachy et al., without determinate phase information, the system is consistent with being in a statistical mixture of interferometer states.

If the system examined by Kovachy et al. does indeed retain quantum coherence over long length and time scales, evidence for such coherence could be obtained either by better phase stabilization of the optical pulses, or, if that is impractical, by operating two well-separated interferometers that share the same optical pulses [9]. Such improvements would allow the impressive technical advance in atom interferometry reported in Ref. [1] to become a test of quantum physics at long length scales.